\documentclass{moriond}

\usepackage{amssymb}
\def\Journal#1#2#3#4{{#1} {\bf #2}, #3 (#4)}

\def\PLB{{\em Phys. Lett.}  B}

\def\PRD{{\em Phys. Rev.} D}

\def\JHEP{\em J. High Energy Phys.}

\def\ra{\rightarrow}

\def\be{\begin{equation}}
\def\ee{\end{equation}}
\def\bea{\begin{eqnarray}}
\def\eea{\end{eqnarray}}

\newcommand{\Photo}{}

\begin{document}
\vspace*{4cm}
\title{$ZH\eta$-vertex: Effective Field Theory Analysis and the Behavior in the Simplest Little Higgs Model}

\author{Ying-nan Mao}
\address{Center for Future High Energy Physics $\&$ Theoretical Physics Division,\\
Institute of High Energy Physics, Chinese Academy of Sciences, Beijing 100049, China}
\author{Chen Zhang}
\address{Physics Division, National Center for Theoretical Sciences, Hsinchu, Taiwan 300}
\author{Shi-Ping He and Shou-hua Zhu}
\address{Institute of Theoretical Physics $\&$ State Key Laboratory of Nuclear Physics and Technology,\\
Peking University, Beijing 100871, China}

\maketitle\abstracts{We re-analyze the $ZH\eta$-vertex with the form $Z_{\mu}(\eta\partial^{\mu}H-H\partial^{\mu}\eta)$, where $H$ 
is the 125 GeV Higgs boson and $\eta$ is an exotic pseudo-axion, based on the effective field theory (EFT) analysis and choose 
the simplest little Higgs (SLH) model as an example. For a pure gauge singlet pseudoscalar $\eta$, after carefully removing all 
off-diagonal two-point transitions, we show that its coefficient $c_{ZH\eta}$ cannot
appear before $\mathcal{O}(\xi^3)$ level, where $\xi$ is the ratio between the electro-weak scale $v$ and a high scale $f$. The same
behavior arises in the simplest little Higgs (SLH) model, which is quite different from the result that has already existed for a long time.}

\section{Introduction}

The discovery of a 125 GeV Higgs boson (denoted as $H$) \cite{Higgs} indicates the success of the standard model (SM).
However, we usually believe that the SM itself is not the end of the theory, because of some unsolved problems, such as the
hierarchy problem, brayogenesis, or dark-matter origin, etc. A tremendous amount of models have been built to solve these
problems. In most extensions of the SM, the scalar sector is also enlarged, for example, there may exist a pseudoscalar
(denoted as $\eta$). In general, $\eta$ and $H$ can interact with $Z$ boson in the anti-symmetric form $Z_{\mu}(\eta\partial^{\mu}H-
H\partial^{\mu}\eta)$. This vertex can lead to new collider signatures, like the associated production of two scalars or the
cascade decay of the heavier scalar \cite{newph}.

The little Higgs (LH) framework, including several models, were built to solve the little hierarchy problem \cite{LH}.
Among those models, the simplest little Higgs (SLH)
model \cite{SLH} has the minimal extended scalar sector, in which the only additional
scalar is a pseudo-axion $\eta$. We re-analyzed the $ZH\eta$ vertex in this model and found that it should appear at
$\mathcal{O}(\xi^3)$ level \cite{ori} (where $\xi\equiv v/f$, $v=246~\textrm{GeV}$ is the electro-weak scale and $f$ is the breaking scale
of a global symmetry) instead of $\mathcal{O}(\xi)$ level \cite{SLH2} which has already existed for long.


\section{EFT Analysis on $ZH\eta$-vertex}

In general, we consider the effective field theory (EFT) at electro-weak scale $v$ \cite{ori}. At scale $v$, we assume the only exotic
particle is a pseudo-axion $\eta$, which is a pure gauge singlet. To dimension-5 and -6, consider the gauge and CP invariant
operators with also $\eta$ shift symmetry, we have
\be
\mathcal{O}_1=\textrm{i}(\partial^{\mu}\eta)\left(\phi^{\dag}D_{\mu}\phi\right)+\textrm{H.c.},\quad\quad\mathrm{and}\quad\quad
\mathcal{O}_2=\left(\phi^{\dag}D_{\mu}\phi\right)\left((D^{\mu}\phi)^{\dag}\phi\right);
\ee
which are possible to contribute to $ZH\eta$-vertex. Here $\phi\equiv\left((v+H-\textrm{i}\chi)/\sqrt{2},G^-\right)^T$ is the
Higgs doublet in the SM and $D_{\mu}$ is the SM covariant derivation.

Consider the contribution from $\mathcal{O}_1$ and define $\xi\equiv v/f$ as above where $f$ is a high scale, the Lagrangian can be expanded as
\bea
\mathcal{L}&=&\mathcal{L}_{\mathrm{SM}}+\frac{c_1}{f}\mathcal{O}_1\supset
\left(D\phi\right)^2+\frac{1}{2}(\partial\eta)^2+\frac{c_1}{f}\mathcal{O}_1
\supset\frac{1}{2}\left((\partial H)^2+(\partial\chi)^2+(\partial\eta)^2\right)+c_1\xi(\partial_{\mu}\eta)(\partial^{\mu}\chi)\nonumber\\
\label{eft}
&&-m_ZZ_{\mu}\partial^{\mu}(\chi+c_1\xi\eta)
+\frac{g}{2c_W}Z_{\mu}(\chi\partial^{\mu}H-H\partial^{\mu}\chi)-\frac{g}{c_W}c_1\xi HZ_{\mu}\partial^{\mu}\eta.
\eea
In general, we can parameterize the $ZH\eta$-vertex as $\mathcal{L}\supset c_{ZH\eta}Z_{\mu}(\eta\partial^{\mu}H-H\partial^{\mu}\eta)$.
We can extract the anti-symmetric part from the last term in Eq.~\ref{eft}, and it naively shows a $ZH\eta$-vertex at $\mathcal{O}(\xi)$ level.
However, there are still unexpected two-point transitions left. The cross term between $\eta$ and $\chi$
means that further diagonalization in CP-odd scalar part is required, while the vector-scalar transition implies that
$\chi$ is not the exact Goldstone field eaten by $Z$ boson. To the leading order of $\xi$, we can perform the field redefinition
\be
\label{red}
\tilde{\chi}=\left(\chi+c_1\xi\eta\right)\left(1+\mathcal{O}(\xi^2)\right),\quad\quad\mathrm{and}\quad\quad
\tilde{\eta}=\eta\left(1+\mathcal{O}(\xi^2)\right),
\ee
to remove these two-point transitions. Here $\tilde{\chi}$ is corresponding Goldstone field of $Z$ boson. After this procedure,
we show that the last two terms in Eq.~\ref{eft} becomes
\bea
&&\frac{g}{2c_W}Z_{\mu}(\chi\partial^{\mu}H-H\partial^{\mu}\chi)-\frac{g}{c_W}c_1\xi HZ_{\mu}\partial^{\mu}\eta\nonumber\\
&\ra&\frac{g}{2c_W}Z_{\mu}\left((\tilde{\chi}\partial^{\mu}H-H\partial^{\mu}\tilde{\chi})-c_1\xi(H\partial^{\mu}\tilde{\eta}
+\tilde{\eta}\partial^{\mu}H)\right),
\eea
which means $c_{ZH\eta}$ cannot survive at $\mathcal{O}(\xi)$ level. It may appear at $\mathcal{O}(\xi^3)$ or higher level.

The operator $\mathcal{O}_2$ can appear in the Lagrangian as $c_2\mathcal{O}_2/f^2$. This operator does not explicitly contain $\eta$.
However, it contribute an additional $Z_{\mu}(\chi\partial^{\mu}H-H\partial^{\mu}\chi)$ term with the coefficient $gc_2\xi^2/(4c_W)$.
Thus when the operator $\mathcal{O}_1$ also appears, the field redefinition $\chi\ra\tilde{\chi}$ in Eq.~\ref{red}
can contribute to $c_{ZH\eta}$ as $gc_1c_2\xi^3/(4c_W)\sim\mathcal{O}(\xi^3)$. Operators with higher dimension also cannot contribute to
$c_{ZH\eta}$ before $\mathcal{O}(\xi^3)$.

In summary, for a pure gauge singlet pseudoscalar $\eta$, $c_{ZH\eta}$ can be induced at $\mathcal{O}(\xi^3)$ or higher level.
For example, in the $(\mathrm{SU}(3)\times\mathrm{U}(1)/\mathrm{SU}(2)\times\mathrm{U}(1))^2$ LH model \cite{SLH} (known as the
SLH model), $c_{ZH\eta}$ appears at $\mathcal{O}(\xi^3)$ level \cite{ori} as expected. While in the $(\mathrm{SU}(4)/\mathrm{SU}(3))^4$
LH model \cite{SLH}, $c_{ZH\eta}$ appears at $\mathcal{O}(\xi)$ level \cite{SLH2} because $\eta$ mixes with the gauge doublet
component at $\mathcal{O}(\xi)$ level.

\section{An Example: $ZH\eta$-vertex in the SLH Model}

\subsection{Model Construction and Properties}

The SLH is based on a global symmetry breaking patten $(\mathrm{SU}(3)\times\mathrm{U}(1))^2\ra(\mathrm{SU}(2)\times\mathrm{U}(1))^2$ at
a high scale $f$ and thus ten Nambu-Goldstone bosons are generated. The gauge group is also enlarged to $\mathrm{SU}(3)\times\mathrm{U}(1)_X$
which means there exist eight massive gauge bosons. Thus eight of the Nambu-Goldstone bosons are eaten by massive gauge bosons and two are left 
as physical scalars. Two scalar triplets are nonlinear realized as \cite{SLH3}
\be
\Phi_1=\mathrm{e}^{\mathrm{i}\Theta'}\mathrm{e}^{\mathrm{i}t_{\beta}\Theta}\left(\begin{array}{c}\mathbf{0}_{1\times2}\\fc_{\beta}\end{array}\right),
\quad\quad\mathrm{and}\quad\quad
\Phi_2=\mathrm{e}^{\mathrm{i}\Theta'}\mathrm{e}^{-\mathrm{i}\frac{\Theta}{t_{\beta}}}\left(\begin{array}{c}\mathbf{0}_{1\times2}\\fs_{\beta}\end{array}\right).
\ee
Here we define $s_{\beta}\equiv\sin\beta$, $c_{\beta}\equiv\cos\beta$, and $t_{\beta}\equiv\tan\beta$, where $\beta$ is a mixing angle between
two triplets. The matrix fields are defined as
\be
\Theta\equiv\frac{1}{f}\left(\frac{\eta\mathbf{I}_{3\times3}}{\sqrt{2}}+\left(\begin{array}{cc}\mathbf{0}_{2\times2}&\phi\\ \phi^{\dag}&0\end{array}\right)\right),\quad\mathrm{and}\quad\Theta'\equiv\frac{1}{f}\left(\frac{\zeta\mathbf{I}_{3\times3}}{\sqrt{2}}
+\left(\begin{array}{cc}\mathbf{0}_{2\times2}&\varphi\\ \varphi^{\dag}&0\end{array}\right)\right);
\ee
where $\phi$ is the usual Higgs doublet defined as above, $\varphi\equiv((\sigma-\mathrm{i}\omega)/\sqrt{2},x^-)^T$ and $\zeta$ are expected to
be the corresponding Goldstone fields of heavy gauge bosons. All fermion doublets should be enlarged to triplets as well.

The gauge kinetic term should be $(D\Phi_1)^2+(D\Phi_2)^2$ where $D_{\mu}\equiv\partial_{\mu}-\mathrm{i}g\mathbb{V}_{\mu}$ with \cite{SLH3}
\be
\mathbb{V}_{\mu}\equiv\left(\begin{array}{ccc}
\frac{1}{2c_W}Z+\frac{(1-3s_X^2)}{2\sqrt{3}c_X}Z'&\frac{1}{\sqrt{2}}W^+&\frac{1}{\sqrt{2}}Y^0\\
\frac{1}{\sqrt{2}}W^-&-s_WA-\frac{c_{2W}}{2c_W}Z+\frac{(1-3s_X^2)}{2\sqrt{3}c_X}Z'&\frac{1}{\sqrt{2}}X^-\\
\frac{1}{\sqrt{2}}\bar{Y}^0&\frac{1}{\sqrt{2}}X^+&-\frac{1}{\sqrt{3}c_X}Z'
\end{array}\right)_{\mu},
\ee
where $Y^0(\bar{Y}^0)\equiv(Y^1\pm\mathrm{i}Y^2)/\sqrt{2}$ and $\theta_X\equiv\arcsin(t_W/\sqrt{3})$.
The heavy gauge bosons can acquire there massses before the electro-weak symmetry breaking (EWSB) as
$m_X=m_Y=gf/\sqrt{2}$ and $m_{Z'}=\sqrt{2/3}gf/c_X$. Loop corrections can generate EWSB which gives the
masses of $W^{\pm}$, $Z$ and $H$ \cite{SLH}. EWSB also induces further mixing between neutral massive gauge
bosons and we denote the mass matrix as $\mathbb{M}^2_V$ in the basis $(Z,Z',Y^2)$. $\eta$ remains massless
because of an accidental global $\mathrm{U}(1)$-symmetry. If we add a soft $\mathrm{U}(1)$-breaking term
(or called $\mu$-term) as $\mu^2\Phi_1^{\dag}\Phi_2+\textrm{H.c.}$, $\eta$ can acquire
its mass $m^2_{\eta}\approx2\mu^2/s_{2\beta}$. The $\mu$-term can also contribute to EWSB.

Assuming $\beta\geq\pi/4$, direct dilepton resonance search at LHC \cite{dilepton} gives $f\gtrsim7.5~\textrm{TeV}$
\cite{maocpv} thus $\xi\equiv v/f\lesssim0.03\ll1$. We can also obtain $f\lesssim85~\textrm{TeV}$ and $t_{\beta}\lesssim8.9$
from the Goldstone scattering unitarity condition \cite{slhcons}. The EWSB condition requires $m_{\eta}\lesssim1.5~\mathrm{TeV}$
as well \cite{slhcons}.

\subsection{Diagonalization of Scalar Sector and Cancelation of the Two-point Transitions}

After expanding the $(D\Phi_1)^2+(D\Phi_2)^2$ term, the Lagrangian contains
\be
\label{la}
\mathcal{L}\supset\frac{1}{2}\mathbb{K}_{ij}\left(\partial_{\mu}G_i\right)\left(\partial^{\mu}G_j\right)+\mathbb{F}_{pi}V^{\mu}_p\partial_{\mu}G_i
+\frac{1}{2}\left(\mathbb{M}^2_V\right)_{pq}V_p^{\mu}V_{q,\mu}.
\ee
Here $G_i$ runs over $(\eta,\zeta,\chi,\omega)$ as a pseudoscalar and $V_p$ runs over $(Z,Z',Y^2)$ as a neutral massive gauge boson.
After EWSB, we have $\mathbb{K}\neq\mathbf{I}_{4\times4}$ which leads to cross terms in scalar kinetic part. The second term in 
Eq.~\ref{la} should be canceled through adding a gauge-fixing term.

We should choose another basis $S_i$ which is canonically-normalized. Define the inner product $\langle S_i|S_j\rangle\equiv\delta_{ij}$
in the space spanned by $S_i$ (or $G_i$), we can obtain
\be
\langle G_i|G_j\rangle=\left(\mathbb{K}^{-1}\right)_{ij}.
\ee
With this relation, choose a basis $\bar{G}_p=\mathbb{F}_{pi}G_i$, we have
\be
\langle\eta|\eta\rangle=(\mathbb{K}^{-1})_{11},\quad\quad\langle\eta|\bar{G}_p\rangle=0,\quad\quad\textrm{and}\quad\quad
\langle\bar{G}_p|\bar{G}_q\rangle=\left(\mathbb{M}^2_V\right)_{pq}.
\ee
Here $\mathbb{M}^2_V$ is just the neutral gauge boson mass matrix and can be diagonalized through an orthogonal matrix $\mathbb{R}$
as $\left(\mathbb{R}\mathbb{M}^2_V\mathbb{R}^T\right)=m^2_p\delta_{pq}$. We thus can obtain a canonically normalized basis
\be
\label{field}
\left(\tilde{\eta},\tilde{G}_p\right)=\left(\frac{\eta}{\sqrt{(\mathbb{K}^{-1})_{11}}},\frac{1}{m_p}\mathbb{R}_{pq}\mathbb{F}_{qi}G_i\right).
\ee
After choosing the gauge-fixing term
\be
\mathcal{L}_{\mathrm{G.F.}}=-\sum_p\frac{1}{2\xi_p}\left(\partial_{\mu}\tilde{V}^{\mu}_p
-\xi_pm_p\tilde{G}_p\right)^2
\ee
where $\tilde{V}_p$ denotes the mass eigenstate of a massive gauge boson, it is straightforward
to show that $\tilde{G}_p$ is exactly the corresponding Goldstone field of $\tilde{V}_p$ and the mass
of $\tilde{G}_p$ is $\sqrt{\xi_p}m_p$. After the procedures above, all off-diagonal two-point transitions are canceled.

Divide $\mathbb{F}_{4\times3}\equiv\left(\tilde{f}_{1\times3},\tilde{\mathbb{F}}_{3\times3}\right)$ with the vector component
$\tilde{f}_q\equiv\mathbb{F}_{q\eta}$, $\tilde{G}_p$ in Eq.~\ref{field} can be re-expressed as $\tilde{G}_p=R_{pq}\left(\tilde{f}_q\eta
+\tilde{\mathbb{F}}_{qi}G_i\right)/m_p$, or equivalently we have
\be
\label{field2}
G_i=\left(\tilde{\mathbb{F}}^{-1}\mathbb{R}^T\right)_{iq}m_q\tilde{G}_q-\left(\tilde{\mathbb{F}}^{-1}\tilde{f}\right)_i
\sqrt{\left(\mathbb{K}^{-1}\right)_{11}}\tilde{\eta}.
\ee
Eq.~\ref{field2} means the original Goldstone degrees of freedom contain physical $\eta$ field thus $VHG_i$-vertices can also contribute
to the $ZH\eta$-vertex, just like the EFT analysis above.

\subsection{Result of the $ZH\eta$-vertex}

We can parameterize the $VHG_i$-vertices as $\mathcal{L}\supset\mathbb{C}_{pi}V_p^{\mu}(G_i\partial^{\mu}H-H\partial^{\mu}G_i)$ where
$\mathbb{C}_{pi}$ is an element of the $4\times3$ matrix $\mathbb{C}$. Naively we have the coefficient of $c_{ZH\eta}=\mathbb{C}_{Z\eta}
=-g\xi/(\sqrt{2}c_Wt_{2\beta})\sim\mathcal{O}(\xi)$ which has existed for long \cite{SLH2}. However, to obtain the physical vertex, we
first need the diagonalization procedure above.

The physical $ZH\eta$-vertex should be parameterized as $\mathcal{L}\supset
\tilde{c}_{ZH\eta}\tilde{Z}^{\mu}(\tilde{\eta}\partial_{\mu}H-H\partial_{\mu}\tilde{\eta})$. Define
\be
\Upsilon\equiv\sqrt{\left(\mathbb{K}^{-1}\right)_{11}}\left(\begin{array}{c}1\\-\tilde{\mathbb{F}}^{-1}\tilde{f}\end{array}\right)
\ee
which is a $1\times4$ vector to express the $\tilde{\eta}$ component in the fields $(\eta,\zeta,\chi,\omega)$. To the leading order
of $\xi$, it is straightforward to obtain
\be
\label{final}
\tilde{c}_{ZH\eta}=\left(\mathbb{R}\mathbb{C}\Upsilon\right)_Z=-\frac{g\xi^3}{4\sqrt{2}c^3_Wt_{2\beta}}\sim\mathcal{O}(\xi^3).
\ee
For the details of the matrixes, please see the appendix. The final result Eq.~\ref{final} shows that the $ZH\eta$-vertex
can appear at $\mathcal{O}(\xi^3)$ level, just like the behavior of EFT analysis above. The cancelation at
$\mathcal{O}(\xi)$ level arises because $\tilde{\chi}=\left(\chi+\mathcal{O}(\xi)\eta\right)\left(1+\mathcal{O}(\xi^2)\right)$
also holds for the SLH model. The mixing between $Z'$, $Y^2$ and $Z$ also contribute to $\tilde{c}_{ZH\eta}$ at $\mathcal{O}(\xi^3)$ level
corresponding to the $\mathcal{O}_2$ contribution in the EFT analysis.

\section{Conclusions and Discussions}

In this paper, we re-analyze the $ZH\eta$-vertex, $Z_{\mu}(\eta\partial^{\mu}H-H\partial^{\mu}\eta)$, based on EFT formalism. 
If the pseudo-axion $\eta$ is a pure SM gauge singlet, this vertex cannot arise until $\mathcal{O}(\xi^3)$ level. As an example, 
we choose the SLH model and calculate $\tilde{c}_{ZH\eta}$, which is the coefficient of
$ZH\eta$ vertex. To the leading order of $\xi$, we obtain $\tilde{c}_{ZH\eta}=-g\xi^3/(4\sqrt{2}c^3_Wt_{2\beta})\sim\mathcal{O}(\xi^3)$,
which satisfies the EFT analysis, but differs from the result which has already existed for a long time.

Because of the non-canonically normalized scalar kinetic part, there remain unexpected two-point transitions. We provide the standard
diagonalization procedure to remove these transitions, after which the contributions to $\tilde{c}_{ZH\eta}$ exactly cancels with each
other at $\mathcal{O}(\xi)$ level. That is the origin of the difference between the result in this paper and that in previous papers.
That also implies the importance to check the normalization in other models, especially in nonlinear realized composite models.

The result $\tilde{c}_{ZH\eta}\sim\mathcal{O}(\xi^3)$ implies the difficulties to test this vertex at future colliders directly through
$H\eta$ associated production or $H(\eta)$ cascade decay channels, because of the highly suppression $\xi^3<3\times10^{-5}\ll1$.
Besides this, the standard diagonalization procedure would also modify the Yukawa interactions including $\eta$. For example, the
$\eta f\bar{f}$-vertices vanish for $f=u,c,b,\ell$ at tree level \cite{maocpv}, which is also different from previous results
\cite{SLH2,SLH3}. Thus the phenomenology of $\eta$ should be completely re-considered in the SLH and similar models.

\section*{Acknowledgments}

This work was supported in part by the Natural Science Foundation of China (Grants No. 11135003, No. 11375014 and No. 11635001) and the China
Postdoctoral Science Foundation (Grant No. 2017M610992). Shou-hua Zhu was also supported by Collaborative Innovation Center of Quantum Matter
and Center for High Energy Physics at Peking University.

\section*{Appendix}
\label{app}
Here we list some details up to $\mathcal{O}(\xi^3)$ during the calculation above. 

The scalar kinetic matrix
\be
\mathbb{K}=\left(\begin{array}{cccc}1&0&\frac{\sqrt{2}\xi}{t_{2\beta}}-\frac{7c_{2\beta}+c_{6\beta}}{6\sqrt{2}s^3_{2\beta}}\xi^3&-\sqrt{2}\xi+\frac{5+3c_{4\beta}}{3\sqrt{2}s^2_{2\beta}}\xi^3\\
0&1&-\frac{\xi}{\sqrt{2}}+\frac{5+3c_{4\beta}}{12\sqrt{2}s^2_{2\beta}}\xi^3&-\frac{2\sqrt{2}\xi^3}{3t_{2\beta}}\\
\frac{\sqrt{2}\xi}{t_{2\beta}}-\frac{7c_{2\beta}+c_{6\beta}}{6\sqrt{2}s^3_{2\beta}}\xi^3&-\frac{\xi}{\sqrt{2}}+\frac{5+3c_{4\beta}}{12\sqrt{2}s^2_{2\beta}}\xi^3&
1-\frac{5+3c_{4\beta}}{12s^2_{2\beta}}\xi^2&\frac{2\xi^2}{3t_{2\beta}}\\
-\sqrt{2}\xi+\frac{5+3c_{4\beta}}{3\sqrt{2}s^2_{2\beta}}\xi^3&-\frac{2\sqrt{2}\xi^3}{3t_{2\beta}}&\frac{2\xi^2}{3t_{2\beta}}&1
\end{array}\right),
\ee
and thus we have $(\mathbb{K}^{-1})_{11}=1+2\xi^2/s^2_{2\beta}$.

The vector-scalar transition matrix $\mathbb{F}\equiv(\tilde{f},\tilde{\mathbb{F}})$ with
\bea
\tilde{f}&=&gf\left(\frac{1}{\sqrt{2}c_Wt_{2\beta}}\xi^2,\frac{\rho}{t_{2\beta}}\xi^2,
-\xi+\frac{5+3c_{4\beta}}{6s^2_{2\beta}}\xi^3\right)^T;\\
\tilde{\mathbb{F}}&=&gf\left(\begin{array}{ccc}-\frac{\xi^2}{2\sqrt{2}c_W}&\frac{\xi}{2c_W}-\frac{(5+3c_{4\beta})\xi^3}{24c_Ws^2_{2\beta}}&
\frac{\xi^3}{3c_Wt_{2\beta}}\\
\frac{1}{\rho}-\frac{\kappa(1+c_{2W})\xi^2}{\sqrt{2}c_{2W}}&\kappa\xi
-\frac{\kappa(5+3c_{4\beta})\xi^3}{12s^2_{2\beta}}&-\frac{2\kappa\xi^3}{3c_{2W}t_{2\beta}}\\
\frac{-2\xi^3}{3t_{2\beta}}&\frac{\sqrt{2}\xi^2}{3t_{2\beta}}&\frac{1}{\sqrt{2}}\end{array}\right).
\eea
Here $\rho\equiv\sqrt{3/2}c_X$ and $\kappa\equiv(1-3s_X^2)/(2\sqrt{3}c_X)$.

The vector
\be
\Upsilon=c^{-1}_{\gamma+\delta}\left(\begin{array}{c}1\\s^2_{\gamma}t^{-1}_{\beta}-s^2_{\delta}t_{\beta}\\
(c_{2\delta}t_{\beta}-c_{2\gamma}t^{-1}_{\beta})\frac{\xi}{\sqrt{2}}\\ \frac{1}{2}(s_{2\delta}t_{\beta}+s_{2\gamma}t^{-1}_{\beta})\end{array}\right)
=\left(\begin{array}{c}1+\frac{\xi^2}{s^2_{2\beta}}\\-\frac{\xi^2}{t_{2\beta}}\\-\frac{\sqrt{2}\xi}{t_{2\beta}}
-\frac{3-c_{4\beta}}{\sqrt{2}t_{2\beta}s^2_{2\beta}}\xi^3\\ \sqrt{2}\xi+\frac{3-c_{4\beta}}{3\sqrt{2}s^2_{2\beta}}\xi^3\end{array}\right);
\ee
where $\gamma\equiv\xi t_{\beta}/\sqrt{2}$ and $\delta\equiv\xi/(\sqrt{2}t_{\beta})$.

The $VHG_i$-vertices matrix is
\be
\mathbb{C}=g\left(\begin{array}{cccc}-\frac{\xi}{\sqrt{2}c_Wt_{2\beta}}+\frac{(7c_{2\beta}+c_{6\beta})\xi^3}{6\sqrt{2}c_Ws^3_{2\beta}}&
\frac{\xi}{2\sqrt{2}c_W}-\frac{(5+3c_{4\beta})\xi^3}{12\sqrt{2}c_Ws^2_{2\beta}}&
-\frac{1}{2c_W}+\frac{(5+3c_{4\beta})\xi^2}{12c_Ws^2_{2\beta}}&-\frac{\xi^2}{2c_Wt_{2\beta}}\\
-\frac{\rho\xi}{t_{2\beta}}+\frac{\rho(7c_{2\beta}+c_{6\beta})\xi^3}{6s^3_{2\beta}}&
\frac{\rho\xi}{2}-\frac{\rho(5+3c_{4\beta})\xi^3}{12s^2_{2\beta}}&
-\kappa+\frac{\kappa(5+3c_{4\beta})\xi^2}{6s^2_{2\beta}}&\frac{\kappa\xi^2}{c_{2W}t_{2\beta}}\\
\frac{1}{2}-\frac{(5+3c_{4\beta})\xi^2}{4s^2_{2\beta}}&\frac{\xi^2}{t_{2\beta}}&
-\frac{\xi}{\sqrt{2}t_{2\beta}}+\frac{(7c_{2\beta}+c_{6\beta})\xi^3}{12\sqrt{2}s^3_{2\beta}}&0\end{array}\right).
\ee

Last, we have the massive gauge boson mixing matrix as
\be
\mathbb{R}=\left(\begin{array}{ccc}1&-\frac{\kappa\rho^2\xi^2}{2c_W}&-\frac{\sqrt{2}\xi^3}{3c_Wt_{2\beta}}\\
\frac{\kappa\rho^2\xi^2}{2c_W}&1&-\frac{2\sqrt{2}(1+2c_{2W})\kappa\xi^3}{3c_{2W}t_{2\beta}}\\
\frac{\sqrt{2}\xi^3}{3c_Wt_{2\beta}}&\frac{2\sqrt{2}(1+2c_{2W})\kappa\xi^3}{3c_{2W}t_{2\beta}}&1\end{array}\right),
\ee
from which we have that the $Z-Z'$ mixing appears at $\mathcal{O}(\xi^2)$ level and $Z-Y^2$ mixing appears at
$\mathcal{O}(\xi^3)$ level.

\end{document}